# Charm Changing Neutral Currents and Supersymmetry[*]

Kwong Lau[**]

*Physics Department, University of Houston, Houston, Texas*

## Abstract

Flavor changing neutral currents (FCNCs) in the charm system are highly suppressed in the standard model (SM). The theoretical strategies used to suppress FCNCs induced by supersymmetry in the strange and beauty systems need not apply to the charm system. The charm changing neutral current decay $D^0 \to \mu^+\mu^-$ is studied phenomenologically in the framework of supersymmetric extensions of the standard model. It is found that the $D^0 \to \mu^+\mu^-$ decay branching ratio can be enhanced to about $10^{-10}$, by having heavy non-degenerate supersymmetric strange or bottom quarks, with negligible effect on the strange and beauty FCNC processes. The $D^0\overline{D}^0$ mixing rate also receives enhancement in these models, but is generally below the present experimental limit. The prospects for measuring $D^0 \to \mu^+\mu^-$ to $10^{-10}$ levels, a definitive signature of new physics beyond the SM, in the near future are discussed.

## Introduction

Flavor changing neutral current (FCNC) processes, such as the dimuon decay and particle-antiparticle mixing of flavored neutral pseudoscalar mesons ($K_L, D^0$, and $B^0$), have provided critical guidance and constraints in the development of the standard model (SM) in the last few decades [1]. The suppression of the strangeness changing neutral current (SCNC) decay, $K_L \to \mu^+\mu^-$, has historically motivated the GIM mechanism [2] which has since become a property of the three-generation SM. The observed branching ratios (BRs) (or upper limits if the processes are not observed) of FCNC processes for the strange, charm, and beauty systems are listed in Table 1, along with the SM short-distance predictions. The non-observation of the charm changing neutral current (CCNC) decay $D^0 \to \mu^+\mu^-$ [3, 4] and beauty changing neutral current (BCNC) decay $B^0 \to \mu^+\mu^-$ [5] at the present level of sensitivity is consistent with the SM. FCNC processes are induced by loop diagrams involving quarks and weak bosons in the SM; the GIM mechanism which predicts exact cancellation of the loop diagrams is evaded by the non-degenerate quark masses. The short-distance FCNC decay and mixing rates have been calculated in the SM [6]; the results depend on the mass-squared differences among the quarks, as well as the CKM couplings. FCNCs in the charm sector are extremely small as a result of the small strange quark mass ($\leq 0.5$ GeV). The beauty and strange systems, on the other hand, receive large contributions from the heavy top quark (of mass $m_t$=174 GeV) in the loop diagrams, typically accounting for 1 to 10 % of the observed rate.

---

[*] Invited talk at the *Richard Arnowitt Fest*, Texas A&M University, College Station, April 5-8, 1998. Published in Proceedings of the Richard Arnowitt Fest, *Relativity, Particle Physics, and Cosmology*, edited by R. E. Allen, World Scientific, 1999.

[**]Email : Lau@uh.edu. This research was supported in part by DOE under contract DE-FG03-96-ER41004 and in part by the Texas ARP Grant no. 003652-828.

Table 1. Some examples of FCNC processes.

| FCNC Process | Experiment | SM (short-distance) Prediction |
|---|---|---|
| $BR(K_L \to \mu^+\mu^-)$ | $(7.2 \pm 0.5) \times 10^{-9}$ | $\approx 10^{-9}$ |
| $BR(D^0 \to \mu^+\mu^-)$ | $< 4 \times 10^{-6}$ | $\approx 10^{-19}$ |
| $BR(B_d^0 \to \mu^+\mu^-)$ | $< 2.6 \times 10^{-7}$ | $\approx 10^{-10}$ |
| $\Delta m_K$ | $(3.49 \pm 0.01) \times 10^{-15}$ GeV | $\approx 3 \times 10^{-17}$ GeV |
| $\Delta m_D$ | $< 1.4 \times 10^{-13}$ GeV | $< 10^{-16}$ GeV |
| $\Delta m_B$ | $(3.12 \pm 0.20) \times 10^{-13}$ GeV | $\approx 3 \times 10^{-13}$ GeV |

The SM, however, is incomplete. Extensions of the SM usually contain new heavy particles and tree-level FCNC couplings, producing in general unacceptably large FCNC decay and mixing rates in the strange and beauty systems. A primary requirement in building realistic models beyond the SM is to have a natural way to suppress FCNCs. The suppression need not apply to the charm system because the SM prediction is well below the present experimental limit (see Table 1). Supersymmetry (SUSY) is a popular extension of the SM, which predicts supersymmetric partners, differing by 1/2 unit in spin, for every known particle. Since no SUSY particles have been observed to date, SUSY particles are expected to be heavy (>100 GeV). FCNCs induced by SUSY particles in loop diagrams are model-dependent, but generally depend sensitively on the mass spectrum of the supersymmetric partners of the quarks (squarks), which in turn depends on the details of the symmetry breaking. The symmetry breaking mechanism of SUSY is not known, but falls into two main categories : supersymmetry is broken spontaneously (SBSS) or explicitly (EBSS). In SBSS models, SUSY-induced FCNCs can be suppressed by requiring the first two generations of squarks to be degenerate in mass by invoking additional symmetries; the third generation squarks can be non-degenerate as long as their couplings to the first two generations are weaker than those of the CKM matrix. Effectively, all the SUSY particles have a universal mass of the order $m_0$, where $m_0$ is related to the energy scale of the symmetry breaking. The present experimental bounds from direct searches place $m_0$ above 0.1 TeV. $m_0$ is generally expected to be above 1 TeV. In some examples of SBSS [7], the mass-squared difference of the squarks is the same as that of the quarks. That is,

$$\Delta m_{ij}^2 = m_{\tilde{q}_i}^2 - m_{\tilde{q}_j}^2 = m_{q_i}^2 - m_{q_j}^2, \qquad (1)$$

where $m_{\tilde{q}_i}, m_{q_i}$ are the squark and quark masses for the i$^{\text{th}}$ generation, respectively. (The $s$ and $t$ squarks are assumed to be degenerate in mass here.) It has been shown in this case the SUSY-induced FCNC effects are smaller than those predicted in the SM [8]. Such a realization of SUSY, however, is unnatural because the fractional mass-squared difference between the first and second generation down-type squarks, $\varepsilon_{21} = \Delta m_{21}^2 / m_0^2 \approx (m_s^2 - m_d^2)/m_0^2 \leq 2 \times 10^{-9}$, is unnaturally small for models with $m_0 > 1$ TeV. A priori, one expects the squark mass differences to be the same order as the squark masses themselves. For example, theories motivated by supersymmetric unification of gravity (SUGRAs) [9] predict SUSY particles with $\varepsilon_{21} \approx O(0.01-1)$. In such models, CCNCs are enhanced by the factor $\approx (\Delta m_{21}^2 / m_s^2)^2 \approx O(10^{-2\pm 2}) \times (m_0 / m_s)^4$. There are other sources of FCNCs in EBSS models, such as flavor-nonconserving coupling between the neutral SUSY partners of the neutral gauge particles (gauginos) and the squarks, contributing also potentially large effects in FCNCs. Gaugino-induced FCNCs in SUSY models and their suppression have been discussed elsewhere [10], and will not be discussed here.

The $D^0 \overline{D}^0$ mixing amplitude is related to the $D^0 \to \mu^+ \mu^-$ decay amplitude via some common loop diagrams. A non-trivial constraint on models with SUSY enhanced CCNCs is the experimental limit on $D^0 \overline{D}^0$ mixing. The present experimental upper limit on $D^0 \overline{D}^0$ mixing is not very stringent. The mixing parameter, defined as $x_D = \Delta m_D / \Gamma_D$, where $\Delta m_D$ is the mass difference between the two mass eigenstates, is less than 0.09 [5]. SUSY models with large CCNCs have to meet the $D^0 \overline{D}^0$ mixing experimental constraint.

The purpose of this talk is to examine the $D^0 \to \mu^+ \mu^-$ decay rate for a generic SUSY model at a phenomenological level. The squark masses are not given by the model, but are rather treated as free parameters. It is argued that this is a meaningful exercise because the lowest-order squark mass spectrum in any given model is likely to be modified by radiative corrections and other effects to a level that is relevant to this study. The $D^0 \to \mu^+\mu^-$ decay and the $D^0\bar{D}^0$ mixing rates are computed as a function of $\varepsilon_{21} = \Delta m_{21}^2/m_0^2$. It is shown that for a reasonable range of this parameter, the $D^0 \to \mu^+\mu^-$ decay rate can be enhanced by many orders of magnitude to a level which can be detected in the near future while the $D^0\bar{D}^0$ mixing rate, though also enhanced by several orders of magnitude above the SM prediction, stays below the current experimental limit.

## $D^0 \to \mu^+\mu^-$ Decay Rate in Some SUSY Models

The effective FCNC couplings for SBSS models have been calculated by Inami and Lim [8]. With a trivial change of notations, the results of Ref. 8 which were derived for SCNCs, can be transcribed for CCNCs. As a result of the super-GIM mechanism, the total amplitude for the $D^0 \to \mu^+\mu^-$ decay rate can be written as a sum over the second and third generations. Since the quark and squark mass-squared matrices can be diagonalized by the same set of unitary matrices in SBSS models, the SUSY CKM matrix is assumed to be the same as that of the ordinary CKM matrix. The effective CCNC Lagrangian is :

$$L_{eff} = \frac{\alpha}{4\pi\sin^2\theta_W} \frac{G_F}{\sqrt{2}} \left[ 4 \sum_{i\geq 2} U_{ci}^* U_{iu} \tilde{C}(\varepsilon_i, x; \delta_i, y)(\bar{c}_L\gamma_\mu u_L)(\bar{\mu}\gamma^\mu\mu_L) + \right.$$

$$\left. \sum_{i,j\geq 2} U_{ci}^* U_{iu} U_{cj}^* U_{ju} \tilde{E}(\varepsilon_i,\varepsilon_j, x; \delta_i, \delta_j, y)(\bar{c}_L\gamma_\mu u_L)^2 \right],$$

where the coefficients $\tilde{C}$ and $\tilde{E}$ are given by

$$\tilde{C}(\varepsilon_i, x, \delta_i, y) = \frac{1}{2}\left\{ f(\varepsilon_i, x) - f(0, x) - \frac{1}{\eta_1}[g(\varepsilon_i, 0, x) - g(0, 0, x)] - \eta_2\delta_i f(\delta_i, y) \right\}$$

and

$$\tilde{E}(\varepsilon_i,\varepsilon_j, x; \delta_i, \delta_j, y) = -\frac{1}{4\eta_1}\left\{ g(\varepsilon_i,\varepsilon_j, x) - g(\varepsilon_i, 0, x) - g(0,\varepsilon_j, x) + g(0,0,x)) \right\} - \frac{1}{4}\eta_2\delta_i\delta_j g(\delta_i, \delta_j, y),$$

respectively. The *f* and *g* functions are defined by

$$f(\varepsilon, x) = \frac{1}{1-x-\varepsilon} + \frac{x+\varepsilon}{(1-x-\varepsilon)^2} \ln(x+\varepsilon)$$

and

$$g(\varepsilon_i, \varepsilon_j, x) = \frac{1}{\varepsilon_i - \varepsilon_j} \left[ \left( \frac{x+\varepsilon_i}{1-x-\varepsilon_i} \right)^2 \ln(x+\varepsilon_i) - (\varepsilon_i \to \varepsilon_j) \right] + \frac{1}{(1-x-\varepsilon_i)(1-x-\varepsilon_j)}.$$

The dimensionless parameters $x, y, \varepsilon_i, \delta_i, \eta_1,$ and $\eta_2$ are given by

$$x = y = \frac{m_{\tilde{d}}^2}{m_{\tilde{W}}^2}, \quad \varepsilon_i = \delta_i = \frac{m_{\tilde{q}_i}^2 - m_{\tilde{d}}^2}{m_{\tilde{W}}^2}, \quad \text{and} \quad \eta_1 = \eta_2 = \frac{m_{\tilde{W}}^2}{m_W^2}.$$

Using standard notations, the $D^0 \to \mu^+ \mu^-$ decay rate is given by [11]

$$\Gamma(D^0 \to \mu^+ \mu^-) = \frac{G_F^4 \, m_W^4 \, f_D^2 \, m_\mu^2}{32 \pi^3} | \sum_{j=s,b} U_{cj}^* U_{uj} \tilde{C}(\varepsilon_j) |^2,$$

where $f_D \approx 200$ MeV is the $D^0$ decay constant. Using the same notations, the $D^0 \overline{D}^0$ mixing rate is given by

$$\Delta m_D = \frac{G_F^2 \, f_D^2 \, B_D \, m_D \, \eta \, m_W^2}{6 \pi^2} | \text{Re} \sum_{j,k=2} U_{jc}^* U_{ju} U_{kc}^* U_{ku} \tilde{E} |,$$

where $B_D$ and $\eta_D$, both of order unity, are QCD correction factors. It is assumed here that all the SUSY particles have approximately the same mass $m_0$, except for the squarks whose mass-squared differences $\varepsilon_i$ are treated as free parameters. It is noteworthy that the apparent singularity for $x = 1$ in $f$ and $g$ is superficial. The numerical values for $\tilde{C}$ and $\tilde{E}$ are evaluated as a function of $\varepsilon$, for $m_0 = m_W$, 0.1, 1, and 10 TeV. The absolute values for $\tilde{C}$ and $\tilde{E}$ are shown in Figures 1a and 1b, respectively. As one can see, $\tilde{C}$ and $\tilde{E}$ increase as $\varepsilon$ and $\varepsilon^2$, respectively, up to $\varepsilon \approx 1$. The SM prediction [6] is also shown in Figure 1 for comparison (the sign of the SM amplitude is opposite to that of SUSY). The dotted lines in Figure 1 labeled strange quark correspond to the values of $\tilde{C}$ computed for mass-squared difference given by Equation 1 ($\Delta m_{21}^2 = m_{\tilde{s}}^2 - m_{\tilde{d}}^2 = m_s^2 - m_d^2$), for the three different values of $m_0$. The band corresponds to $0.04 < \Delta m_{21}^2 < 0.1$ GeV$^2$, a range based on accepted values for the strange and down quark masses. As one can see, $\tilde{C}$ is very insensitive to the value of $m_0$ if the mass relation (1) is observed. In this case, the SUSY contribution to CCNCs is small compared to that of the SM, a well-known result in Ref. 8. The curve labeled top quark corresponds to a mass-squared difference $\Delta m_{21}^2 = m_{\tilde{s}}^2 - m_{\tilde{d}}^2 = m_t^2$, where $m_t$ is the mass of the top quark. For $\varepsilon$ lying between these two extreme cases, the value of $\tilde{C}$ for $m_0 > 1$ TeV can be several orders of magnitude above the SM prediction. For example, for $\varepsilon_2 \approx 10^{-3}$ and $m_0 = 1$ TeV, typical in models motivated by supersymmetric quark flavor symmetry [12], the $D^0 \to \mu^+\mu^-$ rate can be enhanced by more than six orders of magnitude above that of the SM. Similarly, the numerical values for $\tilde{E}$ are shown in Figure 1b with similar notations. It is worth pointing out that the SUSY amplitude $\tilde{E}$ is always lower than the SM prediction for $\varepsilon < 10^{-4}$.

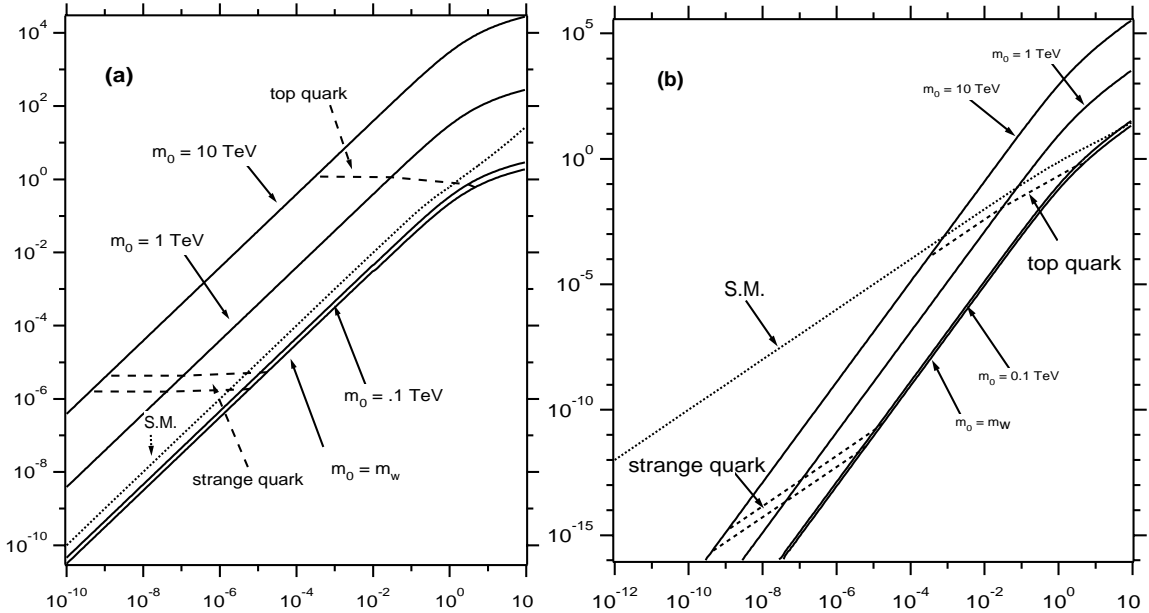

Figure 1. $\tilde{C}$ versus $\varepsilon$ (a) and $\tilde{E}$ versus $\varepsilon$ (b). See text for details.

The $D^0\overline{D}^0$ mixing parameter ($x_D = \Delta m_D / \Gamma_D$) and the $D^0 \to \mu^+\mu^-$ BR, computed according to equations above, are plotted in Figure 2. They are related by the curves for each choice of $m_0$ value, where $\varepsilon$ is the parameter along the curves. The standard CKM matrix values $U_{cs} \approx 0.98$ and $U_{su} \approx 0.22$ are used. A non-trivial observation from Figure 2 is that both the $D^0 \to \mu^+\mu^-$ decay and the $D^0\overline{D}^0$ mixing rates are above the SM predictions if the supersymmetric strange quark is not unnaturally degenerate with the supersymmetric down quark. For example, BR($D^0 \to \mu^+\mu^-$) $\approx 10^{-10}$ and $x_D \approx 10^{-3}$ can be achieved simultaneously with $\varepsilon \approx 10^{-2}$ and $m_0=1$ TeV. Plotted in Figure 2 are also regions excluded by experiment for the $D^0\overline{D}^0$ mixing parameter and the $D^0 \to \mu^+\mu^-$ decay BR. To summarize : if the mass relation (1) is relaxed for the scalar strange quark, there is ample parameter space for SUSY models in which the $D^0 \to \mu^+\mu^-$ decay rate can be many orders of magnitude higher than that of the SM prediction, at a level detectable in the near future, without violating the present experimental bound on $D\bar{D}$ mixing.

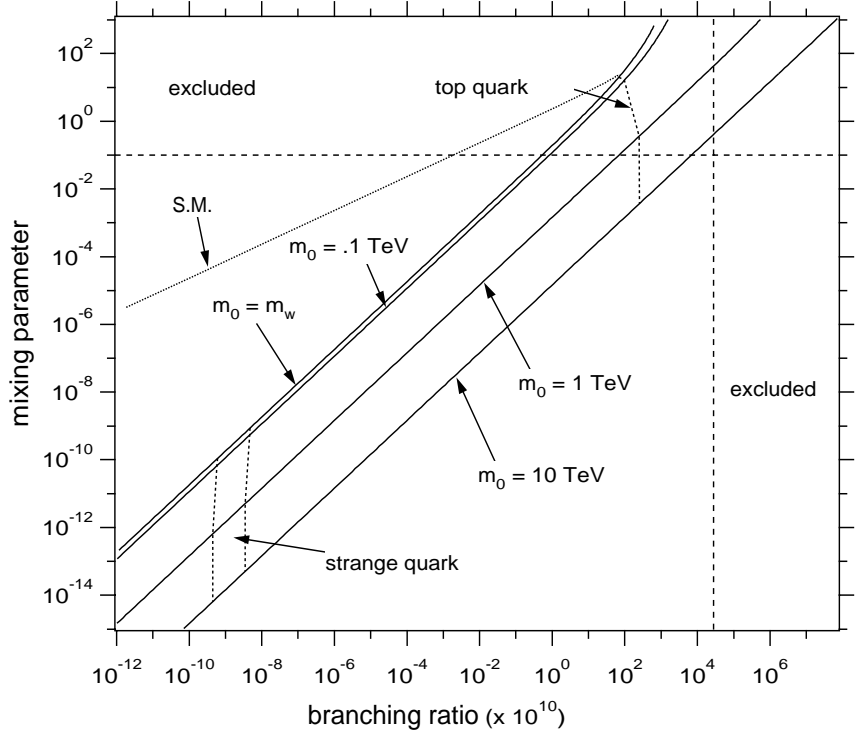

Figure 2. $D^0\bar{D}^0$ mixing rate versus $D^0 \to \mu^+\mu^-$ decay branching ratio for different values of $m_0$. Regions above and to the right of the dotted lines are excluded by experiment.

**Prospects for Detecting $D^0 \to \mu^+\mu^-$ in the Near Future**

Stringent tests of SUSY models can be made in the near future by searching for CCNCs via $D^0 \to \mu^+\mu^-$ decay or $D^0\bar{D}^0$ mixing. CCNCs observed at a level comparable to SCNCs and BCNCs are an unequivocal sign of physics outside the SM, for which SUSY is a leading candidate. Charm particles will be produced in large numbers in several experiments planned for the near future [13]. Among them, HERA-B [14] is scheduled to take data in 1998. HERA-B is a hadron beauty factory capitalizing on the idea of an internal target in the 820 GeV proton ring HERA at DESY. Eight thin wires are introduced inside the beam-pipe, which will interact with the off-momentum protons at a typical rate of 40 MHz. When operated at the design luminosity, the HERA-B Experiment is capable of producing $10^{12}$ charm particles per year. Assuming an overall detection efficiency of the order of 0.1-1% for $D^0 \to \mu^+\mu^-$, a search for $D^0 \to \mu^+\mu^-$ events can be made with a sensitivity of $10^{-9}$ to $10^{-10}$ in BR. The technique of searching for rare charm decays, such as $D^0 \to \mu^+\mu^-$, in hadron environments has been demonstrated in several fixed target experiments [3,4,15]. The dimuon final state is easy to trigger. The precision silicon vertex detectors can be exploited to discriminate the dominant background, muons from hadron decays in flight, by requiring a finite separation of the dimuon vertex from the primary interaction point. The two-body decay kinematics offer additional rejection of background. For example, the E771 experiment at Fermilab has performed a search for this decay in its 1992 data, corresponding to $10^{12}$ interactions with an overall detection efficiency of 0.1%. The dimuon spectrum without vertex requirement is shown in Figure 3a. The mass distribution for the unlike-sign dimuons after the vertex and momentum-balance cuts were applied is shown in Figure 3b. The upper limit on the $D^0 \to \mu^+\mu^-$ BR based on zero candidate in the search region is $4.2 \times 10^{-6}$ at 90% confidence level [3].

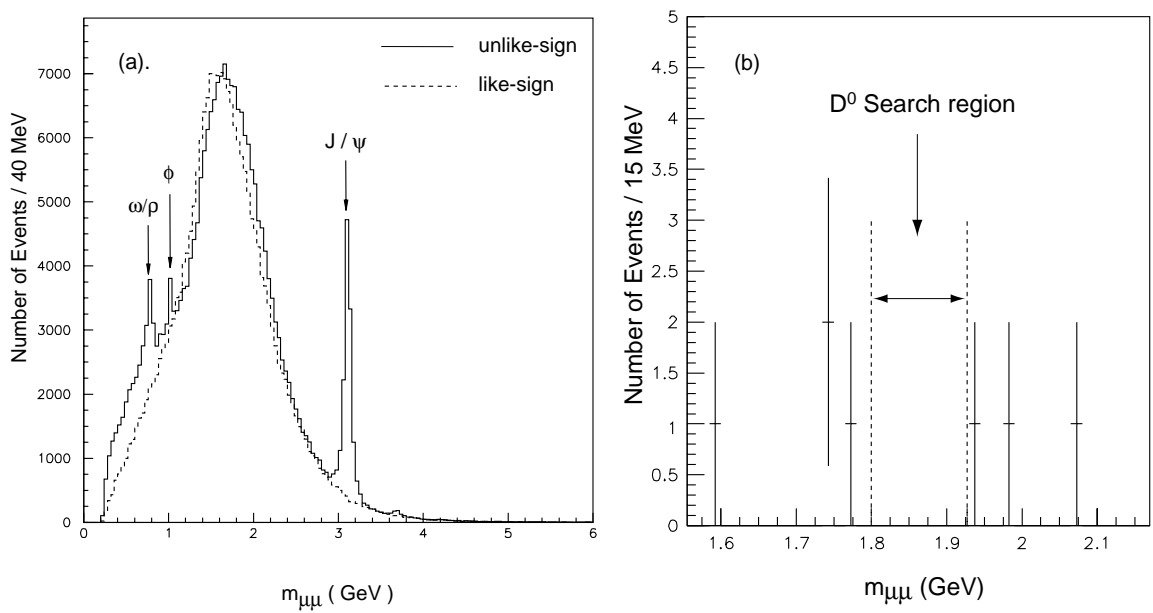

Figure 3. Unlike-sign dimuon spectrum before (a) and after (b) vertex cuts for $D^0 \to \mu^+\mu^-$ search in E771. See Ref. 3 for details of analysis.

The experimental technique for detecting $D^0 \overline{D}^0$ mixing is more involved. Readers are referred to Ref. [16] for details. The search for $D^0 \overline{D}^0$ mixing based on wrong-sign di-lepton signature generally requires a larger sample of $D^0$ mesons. A more sensitive method is to examine the time-dependence of the $D^0 \to K^+ \pi^-$ decay [16].

**Conclusions**

It is shown by numerical calculations that the $D^0 \to \mu^+\mu^-$ BR can be enhanced by many orders of magnitude in some parameter space for SUSY models, to a level detectable in the near future, if the squark mass degeneracy in the charm system is relaxed. This can be achieved without violating the experimental FCNC bounds on the strange and beauty systems. It is noted that the $D^0\overline{D}^0$ mixing rate, while also enhanced by the same mechanism, stays below the current experimental limit. Hence, searching for $D^0 \to \mu^+\mu^-$ in the next generation of heavy quark experiments, such as HERA-B, is a sensitive way to probe SUSY. Unlike direct searches, this method is not limited by the available center-of-mass energy.

**Acknowledgment**

The author thanks the organizers for invitation and hospitality and R. Arnowitt for helpful discussions on SUSY models. The author thanks Keith Decker for his assistance in numerical calculations. This work was supported in part by the DOE (grant no. DE-FG03-96-ER41004) and in part by an ARP grant (project number 003652-828) from the State of Texas.